\documentclass[twocolumn,english]{revtex4}
\usepackage[T1]{fontenc}
\usepackage[latin9]{inputenc}
\usepackage{textcomp}
\usepackage{graphicx}
\usepackage{amssymb}
\usepackage{bm}

\makeatletter

\usepackage{babel}
\makeatother

\newcommand{\units}[2]{#1\nobreak\mbox{$\:\mathrm{#2}$}}

\begin{document}

\title{Wetting on nanorough surfaces}

\author{T. Biben and L. Joly}

\affiliation{Universit\'e de Lyon, F-69000, France; Univ. Lyon 1, Laboratoire PMCN; CNRS, UMR 5586; F-69622 Villeurbanne Cedex}

\begin{abstract}
We present in this letter a free-energy approach to the dynamics of
a fluid near a nano-structured surface. The model accounts both for
the static phase equilibrium in the vicinity of the surface (wetting
angles, Cassie-Wenzel transition) and the dynamical properties like
liquid slippage at the boundary. This method bridges the gap between phenomenological
phase-field approaches and more macroscopic lattice-Boltzmann models.
\end{abstract}

\maketitle

Viscous dissipation is a major problem in micro or even nanofluidic
systems~\cite{Stone2004}. Large pressure gradients are indeed necessary to produce
a significant flow at small scale. This effect is well illustrated
by the Poiseuille law relating the flux to the size of the duct: for
a cylindrical duct of radius $R$, one can check that the flux decreases
like $R^{4}$ when a given pressure gradient is applied. The Poiseuille
law is however obtained by assuming no-slip boundary conditions (BC)
for the liquid at the walls, and recent experiments have shown that this
might not always be the case~\cite{Lauga2005}. In the presence of partial slip BC,
one can expect the situation to be less penalizing, specially if the
slip length associated to the boundary is of the order of the duct
radius or even larger~\cite{Bocquet2007}. Superhydrophobic surfaces have thus attracted
a considerable interest recently as potential candidates for highly
slipping surfaces: if the liquid is repelled by the surface, one can
imagine that a thin layer of gas (air or vapor) at the surface could
produce a lubricating effect. The situation is however more complex
since natural or artificial superhydrophobic surfaces are extremely rough: roughness is indeed a key ingredient
in superhydrophobicity, since it favors the trapping of air or vapor
bubbles at the boundaries~\cite{Lafuma2003}. This effect is due to capillarity, and
the question to understand the coupling between the three ingredients,
hydrodynamics, capillarity and surface roughness is a true challenge.
From the experimental point of view, the situation is still very controversial, with measured slip lengths varying on several orders of magnitude~\cite{Bocquet2007}. It is thus crucial to have numerical models
able to investigate the static and the dynamic properties of a fluid
at a structured boundary. 

Several approaches have been considered (see \cite{Bocquet2007} for a recent review), 
ranging from molecular dynamics at the nanometric scales~\cite{Cottin-Bizonne2003} to hydrodynamic
models (including the lattice-Boltzmann approach~\cite{Dupuis2005}) at micronic scales~\cite{Stroock2002,Cottin-Bizonne2004,Priezjev2005,Sbragaglia2006,Sbragaglia2007,Kunert2007},
and a very recent phase-field model at intermediate scales~\cite{Chakraborty2007}. The method
we shall consider in this letter is intermediate between phase-field
phenomenological approaches, not accounting for the specificities
of the gas phase (low viscosity and high compressibility) and the
more macroscopic lattice Boltzmann approaches. It accounts explicitly for a possible liquid-vapor coexistence
in the vicinity of a structured surface, and also accounts for an hydrodynamic
flow with liquid slippage at the surface. This model is able to reproduce
both static properties such as wetting angles or Cassie-Wenzel states,
but also to predict the effective slip lengths of structured interfaces
by accounting explicitly for the low viscosity gas layer that forms
at the boundary. Moreover, this approach gives access to the intrusion
dynamics of the fluid inside the pores of the structured surface,
which is an important phenomenon when the liquid is submitted to large
pressures.

The model is constructed as follows: 
\begin{equation}
\frac{\partial\rho}{\partial t}+\nabla\cdot(\rho\bm{v})=-C\left(\frac{\delta\Omega}{\delta\rho}\right) \textrm{ or }=C'\Delta\left(\frac{\delta\Omega}{\delta\rho}\right)\label{eq:rho}
\end{equation}
 is the transport equation for the local density field $\rho(\bm{r},t)$,
where $t$ is time and $\bm{r}$ denotes the position; $\bm{v}(\bm{r},t)$
is the local velocity field and $\Omega[\rho]=F[\rho]-\mu\int\rho(\bm{r})\mathrm{d}\bm{r}$ is the grand-potential
functional associated to the thermodynamic equilibrium between the
liquid and the vapor phase ($F[\rho]$ is the free-energy). $\rho$
is thus a physical order parameter here, and not simply a mathematical
object to identify the phases as in usual phase-field approaches \cite{Chakraborty2007}.
 The convective transport of the density field is
included in the left term, while the local thermodynamic equilibrium is 
contained in the right terms: the first writing corresponds to the Allen-Cahn 
model \cite{Cahn77}, interesting for the study of the phase diagram (non-conserved dynamics,
fixed chemical potential $\mu$), and the second writting is the
Cahn-Hilliard model \cite{Cahn61}, corresponding to a more realistic conserved dynamics, that we shall 
use to investigate the non-equilibrium properties. $C'$ is related to the molecular
diffusion constant $D$ in a bulk phase by the relation: $D=C'f"(\rho)$ where $f"$ denotes
the second derivative of the bulk free energy $f(\rho)$ at the bulk density $\rho$. In this work we used
$D=2300\mu m^2 /s$ in the liquid phase, to match the auto-diffusion constant in
pure water ($D=2270\mu m^2 /s$). 
 The free energy
functional is $
F[\rho]=\int \mathrm{d}\bm{r}\left\{ f(\rho)+\frac{w^{2}}{2} |\nabla\rho|^{2}+\rho V_{\text{wall}}(\bm{r})\right\},$
with $f(\rho)=k_{\text{B}}T\rho\left\{ \log\left(\frac{\rho b}{1-\rho b}\right)-1\right\} -a\rho^{2}$
the free energy density of the bulk van der Waals theory ($a$ is the
mean field attractive energy between two atoms, and $b$ is the close
packing inverse density) . The square gradient term accounts for the
natural thickness of the liquid-vapor interface and gives rise to
the surface tension $\gamma_{\text{LV}}=w\int_{\rho_{\text{V}}}^{\rho_{\text{L}}}\sqrt{2(f(\rho)-\mu\rho+P)}\mathrm{d}\rho$
\cite{Rowlinson1989} where $P$ is the bulk pressure, $\rho_{\text{L}}$
and $\rho_{\text{V}}$ are the coexisting liquid and vapor densities respectively.
$w$ is related to the interfacial thickness. $V_{\text{wall}}(\bm{r})$
is the interaction potential applied by the wall. Equation (\ref{eq:rho})
has to be solved with the corresponding equation for the velocity
field:
\begin{equation}
\frac{\partial\rho_{\text{m}}\bm{v}}{\partial t}+\nabla\cdot\left(\rho_{\text{m}}\bm{v}\bm{v}\right)=\nabla\cdot\bm{\sigma}-\rho\nabla\left(\frac{\delta F[\rho]}{\delta\rho}\right)+\bm{f}_{\text{wall}} ,\label{eq:vel}
\end{equation}
where $\rho_{\text{m}}(\bm{r},t)$ is the local mass density ($\rho_{\text{m}}(\bm{r},t)=M\rho(\bm{r},t)$,
with $M$ the molecular mass), $\bm{\sigma} = \eta(\rho)(\nabla\bm{v}+\nabla\bm{v}^{t})$
is the local viscous stress tensor (with $\eta(\rho)$ the shear viscosity).
Although more accurate prescriptions can be considered, we use in
this work the simple ansatz $\eta(\rho)=\rho_{\text{m}}\nu$,
where the kinematic viscosity $\nu$ is assumed to be the same in
the liquid and the vapor phase. The second term $-\rho\nabla\left(\frac{\delta F[\rho]}{\delta\rho}\right)$
is the thermodynamic force field applied by the density field $\rho(\bm{r})$
to the flow. This term contains both the compressibility of the fluid,
fixed by the van der Waals theory, and the capillary force when interfaces
are present. 
The last term of equation (\ref{eq:vel}) accounts
for the interaction with the wall. 

We shall now discuss the interactions between the fluid and the wall.  
The two fields $\rho(\bm{r},t)$ and $\bm{v}(\bm{r},t)$
are defined in the whole parallelepipedic resolution box, and are
thus defined and solved inside the walls. While the static properties of the wall are controlled by $V_{\text{wall}}(\bm{r})$ (expulsion of the fluid from the wall and wetting properties), the boundary conditions for the velocity field are controlled by $\bm{f}_{\text{wall}}$. A prerequisite for a correct description of the dynamical behavior of the liquid at superhydrophobic surfaces is to model properly the intrinsic slip properties of the liquid on a flat solid surface. Our prescription for $\bm{f}_{\text{wall}}$ 
ensures partial slip BC in this case. These boundary conditions correspond to the continuity of the tangential stress $\sigma_{\parallel z}$ at the fluid/solid interface: $\sigma_{\parallel z}$ in the fluid equals $-\lambda v_{\parallel}(z_{\text{s}})$, the friction stress applied by the wall, where $z$ is the normal direction, $\parallel$ the tangential direction under consideration, $\lambda$ is the friction coefficient and $v_{\parallel}(z_{\text{s}})$ the slip velocity at the wall ($z=z_{\text{s}}$). The partial slip BC thus expresses:
\begin{equation}
v_{\parallel}(z_{\text{s}})
= \frac{1}{\lambda}\sigma_{\parallel z} = \frac{\eta}{\lambda} \left.\frac{\partial v_{\parallel}}{\partial z}\right|_{z=z_{\text{s}}}=b\left.\frac{\partial v_{\parallel}}{\partial z}\right|_{z=z_{\text{s}}} ,\label{partial}
\end{equation}
where $b=\eta/\lambda$ is the slip length, and $\eta$ is the shear viscosity of the fluid. If $b$, $\eta$ and the slip velocity are known, the stress at the interface is $-\eta v_{\parallel}(z_{\text{s}})/b$.
\begin{figure}
\includegraphics[width=0.9\columnwidth]{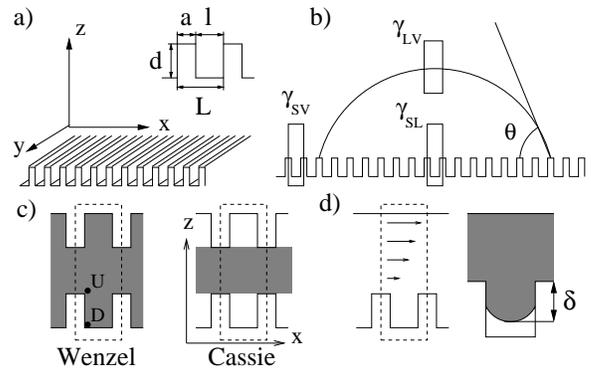}
\caption{\label{cap:dessin}Schematic views of the systems under consideration.}
\end{figure}

The bare interaction between an atom of the wall and an atom of the fluid is the Lennard-Jones potential in our model: $V_{\text{LJ}}(r)=4\epsilon\left[(\sigma/r)^{12}-(\sigma/r)^{6}\right]$. The global potential applied by the wall on a single atom of the fluid is $
V_{\text{wall}}(\bm{r})=\int\rho_{\text{wall}}(\bm{r}')V(|\bm{r}'-\bm{r}|)\mathrm{d}\bm{r}'$,
where $\rho_{\text{wall}}(\bm{r})$ is the density of atoms in the wall
(assumed to be a constant inside the wall and zero elsewhere), and
$V(r)$ is the bare interaction potential [$V_{\text{LJ}}(r)$] multiplied
by the pair distribution function between an atom of the wall and
an atom of the fluid. The
main property of this pair distribution function is to vanish when
$r\to0$ since two atoms cannot overlap, while $V_{\text{LJ}}(r)\to+\infty$
in this limit. The product of these two functions thus goes to a well
defined limit in $r=0$. As a result, the potential energy inside
the wall is large, but finite. We shall denote by $V_{\text{cut}}$ this
energy and use the simple ansatz: $
V_{\text{wall}}(\bm{r})=\min(V_{\text{cut}},\int\rho_{\text{wall}}(\bm{r}')V_{\text{LJ}}(|\bm{r}'-\bm{r}|)\mathrm{d}\bm{r}').$
 This prescription fixes the interaction to be of the Lennard-Jones
type close to the boundary with the fluid, and to be $V_{\text{cut}}$ inside
the wall. $V_{\text{cut}}$ fixes the value of the fluid
density inside the wall, chosen to be negligible as compared to the density of the
vapor phase ($\rho\sigma^{3}\lesssim10^{-3}$). In this limit, the
results become independent of $V_{\text{cut}}$.

The force field $\bm{f}_{\text{wall}}$ exerted by the wall on the fluid is chosen as a friction force:
\[
\bm{f}_{\text{wall}} = - k \epsilon \rho(\bm{r}) \bm{v}(\bm{r}) \int \mathrm{d}\bm{r}' 
\left\{ \rho_{\text{wall}}(\bm{r}') \text{e}^{-\frac{3}{2}\frac{(\bm{r}'-\bm{r})^2}{\sigma^2}}\right\} 
\].
This force is proportional to the interaction energy $\epsilon$ between the solid and the fluid \cite{Barrat1999a,Bocquet2007}. This prescription enabled us to reproduce correctly the evolution of the BC with the wetting properties. In particular, we obtain a transition from a no-slip BC in a wetting situation ($\theta \sim \units{0}{^\circ}$) to a partial slip BC with $b \sim \units{10}{nm}$ (depending on the value of $k$) in a non-wetting situation ($\theta \sim \units{120}{^\circ}$).

The two equations (\ref{eq:rho}) and (\ref{eq:vel}) can be solved
numerically on a regular lattice (cubic unit cell in 3D or square
in 2D) with periodic BC at the edges of the resolution
box on the appropriate quantities, as we shall discuss. The wall is
defined in the resolution box by the function $\rho_{\text{wall}}(\bm{r})$.
We consider a slab geometry corresponding to a fluid confined between
two walls, the bottom wall is textured with a periodic structure (crenels
in this study), while the top wall can either be structured in static
studies (Fig.\ref{cap:dessin}-c), or planar in dynamic situations
(Fig.\ref{cap:dessin}-d). The density $\rho(\bm{r})$ is periodic
in the slab (x-y) direction, but also in the z direction since it
goes to a constant value (nearly vanishing) inside the walls. The
velocity field is periodic in the x-y direction, but not in the z
direction where a linear shear $\bm{v}_{\text{s}}(\bm{r})$
is applied. In this case, the periodic BC is taken
on $\bm{u}(\bm{r}) = \bm{v}(\bm{r})-\bm{v}_{\text{s}}(\bm{r})$
in any directions. 
 By using a simple Euler scheme, equations
(\ref{eq:rho}) and (\ref{eq:vel}) can be solved iteratively starting
from an initial configuration for $\rho(\bm{r})$ and $\bm{v}(\bm{r})$
. Taking advantage of the periodicity of $\rho(\bm{r})$ and $\bm{u}(\bm{r})$,
we use an implicit method in the Fourier space to improve the numerical
stability, but simpler schemes can be used as well.

\begin{figure}
\includegraphics[width=0.9\columnwidth]{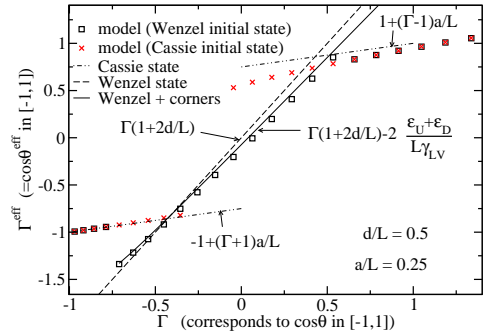}
\caption{\label{cap:Cassie-Wenzel}Wetting properties of a crenelated surface (see text for details).}
\end{figure}

To illustrate the ability of the model to account for the small scale
physics of the interfaces,
 we consider two situations. 
The first study corresponds to a static case, where $\bm{v}_{\text{s}}(\bm{r})=0$.
 The wetting properties
can be probed either by measuring the contact angles of a drop placed
on the surface, or more simply by measuring the three surface tensions
of the liquid-solid ($\gamma_{\text{SL}}$), vapor-solid ($\gamma_{\text{SV}}$) and liquid-vapor ($\gamma_{\text{LV}}$) interfaces. The
model presented here gives a direct access to the free-energy or the
grand-potential (an advantage compared to molecular simulations);
the surface tensions can thus be measured directly by considering
a fluid confined between two structured walls as depicted in Fig.\ref{cap:dessin}-c.
Provided the distance between the walls is large compared to the typical
relaxation length of the density profile, the walls can be considered
as independent. The relaxation length scale of the density profile
is given by the interfacial thickness $w$, and we use a distance
between the two walls of the order of $50w$ , which ensures an accurate
determination of the surface tensions.The liquid-solid [resp. vapor-solid] surface tension
is measured by starting the simulation with a liquid [resp. vapor] phase between
the walls. For planar interfaces this prescription is
very efficient and the two surface tensions can be measured in the
full range of variation of the interaction potential $V_{\text{LJ}}$, which
is done by varying $\epsilon$. From these two quantities, and
the knowledge of $\gamma_{\text{LV}}$, the contact angle with the planar
surface can be calculated through the quantity $\Gamma = \frac{\gamma_{\text{SV}}-\gamma_{\text{SL}}}{\gamma_{\text{LV}}}$.
$\Gamma$ corresponds to $\cos\theta$ when $|\Gamma|\leq1$ and 
is an increasing function of $\epsilon$ (attractions favor wetting
of the liquid phase). We shall consider $\Gamma$ rather than $\epsilon$
in the discussions, since the contact angle is the experimentally
relevant quantity. When the wall is not planar, and has a crenel shape
as depicted in Fig.\ref{cap:dessin}-a, the situation is more complex
since various states can be observed.  We used in this case two different
types of initial configurations: Wenzel states corresponding to a
monophasic system, or Cassie states corresponding to a diphasic situation (see Fig.\ref{cap:dessin}-c). Of course, the equilibrium result
should not depend on the initial state, this prescription gives us
however a way to probe metastability in the system. The wetting properties
of the structured wall can be measured as well by determining the
effective surface tensions $\gamma_{\text{SV}}^{\text{eff}}$ and $\gamma_{\text{LV}}^{\text{eff}}$
for the equilibrated configurations, the variation of the corresponding
wetting parameter $\Gamma^{\text{eff}}$ is plotted in Fig.\ref{cap:Cassie-Wenzel}.
This figure illustrates very well the Cassie-Wenzel transition between the imbibed and super-hydrophobic states~\cite{Lafuma2003}. Moreover, several important features
can be noted from these results: first we observe a strong metastability,
with the presence of superhydrophobic metastable states, and next,
we observe an asymmetry in the wetting properties when we go from
the non-wetting region to the wetting one. The macroscopic model describing
the transition between Cassie and Wenzel states is compared to the
numerical results in Fig.\ref{cap:Cassie-Wenzel}. This model is quite
accurate in the non-wetting region, which is not a surprise since
we have the same geometry: the interfaces are planar at coexistence.
In the wetting regime on the contrary, we observe discrepancies with
the macroscopic theory. The difference comes from corner effects: the interaction
potential between the fluid and the wall has particular values at
the corners that generate a corner energy. These corner effects are
mainly visible when a liquid phase fills the crenels (wetting situation), in gas phases
the low value of the density leads to negligible contributions. 
If $\epsilon_{\text{U}}$ and $\epsilon_{\text{D}}$ are the
excess free-energies in the liquid phase at the upper and the lower corners (see in Fig.1-c the points U and D), the Wenzel theory can be modified as follows:
\begin{equation}
\Gamma^{\text{eff}}=\Gamma(1+2d/L)-2\frac{\epsilon_{\text{U}}+\epsilon_{\text{D}}}{L\gamma_{\text{LV}}}
\end{equation}
where $\epsilon_{\text{U}}$ and $\epsilon_{\text{D}}$ are measured independently. This theory reproduces very well the effective contact angle (Fig.\ref{cap:Cassie-Wenzel}). Please note that $\epsilon_{\text{U}}$ and $\epsilon_{\text{D}}$ are not constant, they vary slightly with $\cos\theta$, which is visible in Fig.\ref{cap:Cassie-Wenzel}.
 
%

We now turn to the dynamical properties of the surfaces. The dynamics of a fluid close to a textured wall
can be accounted for using effective partial slip BC (\ref{partial}),
where both the slip length $b$ and the hydrodynamic position of the wall $z_{\text{s}}$ are unknown. For a smooth surface, it has been shown that the BC applies at about one molecular layer inside the fluid, for both no-slip and slip cases \cite{Barrat1999a}. 
However for superhydrophobic surfaces, it is not clear where the BC should apply. In this work we therefore \emph{measured} both $b$ and $z_{\text{s}}$, by probing our system using both Couette and Poiseuille flows \cite{Barrat1999a}.
\begin{figure}
\includegraphics[width=0.9\columnwidth]{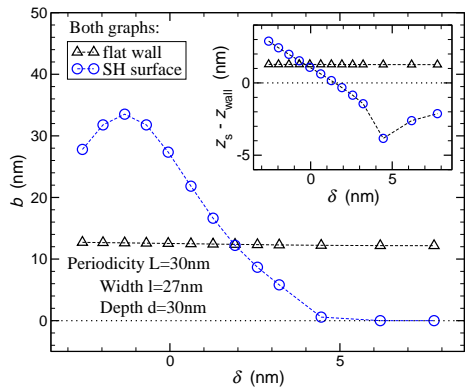}
\caption{Evolution of the slip length $b$ with the recess $\delta$ of the meniscus inside the crenel, for a flow perpendicular to the crenel. Inset: Evolution of the shear plane position $z_{\text{s}}$ as a function of $\delta$. For the SH surface, $z_{\text{w}}$ is defined as the top of the crenels.\label{cap:slip}}
\end{figure}
The geometry of the system corresponds to Fig.\ref{cap:dessin}-d),
where the top wall is planar and the bottom one is structured. This
hybrid geometry allows for the determination of the effective boundaries
for both the planar and the textured interfaces at the same time. 
The figure corresponds to a Couette flow, for which the top wall
moves with a velocity $\bm{U}$ perpendicular to the crenels, while the bottom wall is fixed.
The Poiseuille situation corresponds to fixed walls, but a pressure
gradient is applied. 
We shall investigate the variation of both
$b$ and $z_{\text{s}}$ as a function of the liquid pressure.
Rather than the pressure itself, we can equivalently fix the average
density in the slab by using a conserved dynamics, and measure the
pressure afterwards. 
%
We shall use $\delta$ , the recess length of the liquid in the crenel 
(taken from the top of the crenels, as depicted in Fig.\ref{cap:dessin}-d), as the control parameter. $\delta$
can either be negative or positive.
The slip length of the planar (bare) liquid-wall interface is
taken to be around 13\,nm at coexistence, and is observed to vary very
weakly with pressure (Fig.\ref{cap:slip}). 
On the contrary, the slip length of the structured wall varies
in larger proportions, between 35\,nm for the upper limit down to 0
for the lower one. Pressure has the effect to reduce slippage, and
even to cancel it. The effective slip length crosses the bare
one when $\delta_{\text{c}}\simeq2$\,nm, and is lower above. This result
is largely independent of the crenel depth since the depth we consider
in the present simulation (30\,nm) is large compared to the crossover
length $\delta_{\text{c}}$. 
 
To conclude, the model presented here is very flexible and can account for complex BC.
Although we focussed on crenelated walls, we can treat any type of geometry, 3D shapes such as posts or random surfaces, or conic shapes to model surfaces coated with nanotubes \cite{Joseph2006}. 
The dynamics on crenelated surfaces revealed the strong sensitivity of the effective slip length to the recess of the meniscus $\delta$ inside the crenels, cancelling slippage as soon as $\delta\simeq2$\,nm. This critical value increases with the periodicity $L$, although it remains a small fraction of it. However, increasing $L$ is not necessarily a good idea since the corresponding intrusion pressure will decrease. 
Although nanostructured surfaces have lower slip lengths than microstructured ones in general, the slip properties resist better to pressure. There is thus a compromise to find between slippage and pressure resistance.

\end{document}